\documentclass[preprint2]{aastex6}
\graphicspath{{./}}
\slugcomment{Accepted to Astrophysical Journal, 17 February 2017}

\newcommand{\kms}{km~s$^{-1}$}

\begin{document}

\title{K2 Ultracool Dwarfs Survey I: Photometry of an L Dwarf Superflare}

\author{John E.\ Gizis, Rishi R.\ Paudel}
\affil{Department of Physics and Astronomy, University of Delaware, Newark, DE 19716, USA}
\author{Sarah J.\ Schmidt}
\affil{Leibniz-Institute for Astrophysics Potsdam (AIP), An der Sternwarte 16, 14482, Potsdam, Germany}
\author{Peter K.\ G.\ Williams}
\affil{Harvard-Smithsonian Center for Astrophysics, 60 Garden Street, Cambridge, MA 02138, USA}
\author{Adam J.\ Burgasser}
\affil{Center for Astrophysics and Space Science, University of California San Diego, La Jolla, CA 92093, USA}

\begin{abstract}
We report on K2 Campaign 8 measurements of a huge white light flare on the L1 dwarf SDSSp J005406.55-003101.8 (EPIC 220186653). The source is a typical L1 dwarf at a distance of $\sim50$ pc, probably an old hydrogen-burning star rather than a young brown dwarf.  
In the long (30-minute) cadence photometry, the flare peak is 21 times the flux of the stellar photosphere in the broad optical {\it Kepler} filter, which we estimate corresponds to $\Delta V \approx -7.1$. The total equivalent duration of the flare is 15.4 hr.  We estimate the total bolometric energy of the flare was $4 \times 10^{33}$~erg, more powerful than the previously reported {\it Kepler} white light flares for the L1 dwarf WISEP J190648.47+401106.8, but weaker than the $\Delta V = -11$ L0 dwarf superflare ASASSN-16ae. 
The initial (impulsive) cooling phase is too rapid to resolve with our 30-minute cadence data, but after one hour the gradual cooling phase has an exponential time constant of 1.8 hours.  We use template fitting to estimate that the full-time-width-at-half-amplitude of the light curve is $<10$ minutes and that the true flare maximum reached $\sim70$ times the stellar photosphere, or $\Delta V \approx -8$. This flare is comparable to the most powerful {\it Kepler} flares observed on the active M4 dwarf GJ 1243. 
\end{abstract} 

\keywords{stars: activity --- stars: flare --- stars: chromospheres  --- stars: low-mass  --- stars: individual: SDSSp J005406.55-003101.8}

\section{Introduction\label{intro}}

Building on the long history of ground-based photometry of flares from main sequence stars  (see \citealt{Gershberg:2005lr} for a complete overview), the precision and time coverage of the 
{\it Kepler} mission \citep{2010ApJ...713L..79K} has revitalized the study of white light flares \citep{2011AJ....141...50W}.  \citet{2016ApJ...829...23D} has recently published a catalog of over 850,000 stellar flares detected on 4041 stars during the original {\it Kepler} mission. Important {\it Kepler}-based studies include the characterization of ``super-flares"  (with bolometric energy $>10^{33}$ erg) on solar-type stars \citep{2012Natur.485..478M,2014ApJ...792...67C,2015EP&S...67...59M} and flares on both active and inactive M-dwarf stars \citep{2014ApJ...797..121H}.  Perhaps the best studied flare star with {\it Kepler} is the rapidly rotating, nearby M4 dwarf \object{GJ 1243}, which was monitored for eleven months in short (one-minute) cadence mode, resulting in the detection of numerous well-characterized flares with {\it Kepler}-band energies $10^{29}-10^{33}$ erg \citep{2013MNRAS.434.2451R,2014ApJ...797..122D,2016ApJ...829..129S}. The extended K2 mission \citep{2014PASP..126..398H} allows many new fields to be monitored with the {\it Kepler} photometer, enabling studies of additional nearby M dwarfs  
\citep{2015MNRAS.449.3015R,2016MNRAS.tmp.1060S}.  

White light flares from L dwarfs are rare enough that targeted ground-based photometric campaigns have not detected them \citep{2013MNRAS.428.2824K,2015MNRAS.453.1484R}, but  
\citet{2013ApJ...779..172G} detected 21 white light flares from the nearby L1 dwarf \object{WISEP J190648.47+401106.8} (hereafter W1906+40) in three months of {\it Kepler} short-cadence monitoring. These flares had estimated bolometric energies in the range $6 \times 10^{29}$ to $1.6 \times 10^{32}$ erg.  
A much more powerful L dwarf flare was later discovered during a ground-based supernova search: the flare ASASSN-16ae on the L0 dwarf \object{SDSS J053341.43+001434.1} (hereafter S0533+00) was observed at $\Delta V=-11$ magnitudes \citep{2016ApJ...828L..22S}. Its total estimated bolometric energy of $>6.2 \times 10^{34}$ erg is only a lower limit to the flare energy since the light curve was sparsely sampled.  These flares must be related to chromospheric and coronal activity as in active GKM main sequence stars, yet H$\alpha$ chromospheres and X-ray coronae are known to weaken and disappear for cooler L dwarfs \citep{2000AJ....120.1085G, 2010ApJ...709..332B,2015AJ....149..158S} as the stellar rotation-activity relation breaks down (see \citealt{2014ApJ...785...10C} and references therein.) High-resolution spectroscopy suggests all L0-L1 dwarfs are rapid rotators, implying they do not significantly spin down even over billions of years \citep{2008ApJ...684.1390R}. Some 90\% of L0 dwarfs and 67\% of L1 dwarfs show chromospheric H$\alpha$ emission when observed at sufficiently high resolution and signal-to-noise, but the strength of their
chromospheres as measured by $L_{{\rm H}\alpha}/L_{\rm bol}$ is an order of magnitude weaker than in 
dMe stars, implying early-L dwarfs have cooler chromospheres with lower filling factors \citep{2015AJ....149..158S}. The underlying physical reasons for this decline in traditional activity indicators are thought to be the increasing neutrality and resistance of the outer atmosphere \citep{Mohanty:2002lr}, but additional factors such as changes in the mode of magnetic reconnection \citep{2010ApJ...721.1034M} or dynamo modes \citep{2014ApJ...785...10C} may play important roles. We note as well that the H$\alpha$ emission
of some ultracool dwarfs may be auroral, rather than chromospheric, in nature \citep{2015Natur.523..568H,2016ApJ...826...73P}.

We have been monitoring L dwarfs that lie in each K2 campaign field of view in both long and short cadence mode, with the aim of detecting flares, transits, and cloud variability. Here we report the detection of a white light flare on an L1 dwarf with a total bolometric energy of $4 \times 10^{33}$ erg, and measure the gradual decay phase of an L dwarf flare for the first time.  We present the target and K2 data in Section~\ref{data}, fit model flare templates in Section~\ref{templates}, and discuss the results in Section~\ref{discussion}.  

\section{Target Characteristics and Data}\label{data}

\subsection{Properties of SDSSp J005406.55$-$003101.8}

The discovery of the L1 dwarf \object[SDSS J005406.55-003101.8]{SDSSp J005406.55$-$003101.8} (hereafter S0054$-$00) was first reported by \citet{2002AJ....123..458S} and \citet{2002AJ....123.3409H}, who each presented optical spectra. \citet{2015AJ....149..158S} presented a higher signal-to-noise Baryon Oscillation Spectroscopic Survey (BOSS) optical spectrum which confirms the L1 spectral type. Cross-correlating this spectrum with the \citet{2014PASP..126..642S} L1 dwarf template, we measure $v_{\rm rad}= -14.4 \pm 14.3$ \kms.  We set a 3-sigma limit of 5\AA~on the equivalent width of the H$\alpha$ emission which implies $\log (L_{\rm{H}\alpha}/L_{\rm bol}) < -5.0$.  The \citet{2015AJ....149..158S} survey found the average $\log(L_{\rm{H}\alpha}/L_{\rm bol})= -5.31$ for L1 dwarfs, so S0054$-$00 was not unusually active when the BOSS spectrum was taken, and indeed may have a typical chromospheric activity level.  \citet{2014ApJ...794..143B} obtained a near-infrared spectrum and also classified S0054$-$00 as an L1 dwarf and found no evidence of a cooler unresolved companion in their spectral fitting. We measure the proper motion from a linear fit to the 2MASS \citep{2mass}, SDSS \citep{Aihara:2011fk}, and WISE \citep{2010AJ....140.1868W} positions. From the observed SDSS optical and 2MASS infrared magnitudes,
the \citet{2012ApJS..201...19D} parallax compilation, and the latest color-absolute-magnitude relations (Schmidt et al., in prep.), we estimate a distance of $d=50.9 \pm 9.6$ pc.
The observed and derived properties of the target are given in Table~\ref{tab:prop}. From its colors and spectra, S0054$-$00 appears to be a typical field L1 dwarf, and with a galactic space component motion ${\rm V}=-46 \pm 13$ \kms, it is most likely to be an old (age $>1$ billion years) hydrogen-burning star (see \citealt{2014AJ....147...94D}). From the $V-J$ colors of other L1 dwarfs \citep{dahn,2014AJ....147...94D} we estimate that the apparent magnitude is $V \approx 24.1$.

\begin{deluxetable}{lc} \tablewidth{0pt} \tabletypesize{\scriptsize}
\tablecaption{Properties of S0054$-$00 \label{tab:prop} }
\tablehead{ \colhead{Parameter}  & \colhead{Value}  }
\startdata
\multicolumn{2}{c}{Photometric} \\
\hline
$r$ & $22.50\pm0.15$ \\ 
$i$ & $20.09\pm0.03$ \\
$\widetilde{K}_p $ & $20.5\pm0.1$ \\
$z$ & $18.25\pm0.02$ \\
$J$ & $15.73\pm0.05$ \\
$H$ & $14.89\pm0.05$ \\
$K_S$ & $14.38\pm0.07$ \\
$W1$ & $14.00\pm0.03$ \\
$W2$ & $13.76\pm0.05$ \\
\hline
\multicolumn{2}{c}{Kinematic} \\
\hline
d ($M_i$/$i-K_S$) [pc]\tablenotemark{a}  & 50.9$\pm$9.6 \\
$\mu_{\alpha}$ [mas yr$^{-1}$] & $194.14\pm$18.24 \\
$\mu_{\delta}$ [mas yr$^{-1}$] & $-148.01\pm$37.83 \\
$V_{tan}$ [km s$^{-1}$] & $58.9\pm$14.3 \\ 
$V_{rad}$ [km s$^{-1}$] & $-14.4\pm$14.3 \\
${\rm U}$ [km s$^{-1}$] & 7.1$\pm$8.0 \\ 
${\rm V}$ [km s$^{-1}$] & $-$45.5$\pm$13.4 \\
${\rm W}$ [km s$^{-1}$] & 4.6$\pm$13.7 \\
\enddata
\tablenotetext{a}{Based on Schmidt et al. (2016) in prep}
\end{deluxetable}

\subsection{K2 Photometry}

K2 measured S0054$-$00 as EPIC 220186653 during Campaign 8 (2016 Jan 3  - 2016 Mar 23)
in long cadence (30 minute) mode \citep{Jenkins:2010fk}. There are 3445 good quality measurements over 78.6 days (Kepler mission dates 2559.11 to 2637.75); each data point is the average flux during a 29.4 minute period. We measure aperture photometry from the pixel files using the {\textsf Astropy}-affiliated {\textsf photutils} package. Rather than centroid on the noisy target in each frame, we adopt a best position based on the median of all centroid measurements, and then adjust it for each observation using the spacecraft motion estimate calculated by the mission (recorded as POS\_CORR1 and POSS\_CORR2 in the FITS file headers). 
For the eight hours before the flare, the median count rate is 84.5 count s$^{-1}$ through both 2 and 3 pixel apertures; we adopt the 2 pixel aperture photometry for our analysis. 

The broad {\it Kepler} filter extends from 430nm to 900nm. As discussed by \citet{2013ApJ...779..172G}, an L1 dwarf photosphere contributes significant counts only from the reddest part of this range, but hot flares contribute through the range and therefore have a higher mean energy per observed count.
Because the principal goals of the original {\it Kepler} mission did not require
absolute calibration of the photometry, the mission relied on
$K_p$, an AB-magnitude system that used ground-based $gri$ photometry
to predict what magnitude {\it Kepler} would observe  
\citep{2011AJ....142..112B}. 
$K_p$ values for K2 fields have been determined by \citet{2016ApJS..224....2H}, who remark that the 
predicted magnitudes of red M dwarfs are too bright. We therefore do not expect the
catalog value $K_p=17.2$ for the extremely red S0054$-$00 to be useful. 
\citet{2015ApJ...806...30L}, \citet{2015MNRAS.447.2880A}
and \citet{2016MNRAS.456.1137L} find that for most (AFGK) stars, $K_p$ predicts the 
observed K2 count rate well, with a zero-point of 25.3 for a 3-pixel
aperture.  To clearly distinguish between the ground-based, catalog
$K_p$ value, and the actual space-measured magnitude of S0054$-$00, 
we follow \citet{2015ApJ...806...30L} in defining:

$$\widetilde{K}_p \equiv 25.3 - 2.5 \log({\rm count~rate})$$

Under this system, $\widetilde{K}_p = 20.5$ for our target before the flare. Our
observed signal-to-noise per 30-minute cadence, 13, is consistent with 
other K2 stars \citep{2016MNRAS.456.1137L} of similar magnitude. 
The K2 photometry of our target does show long-term drifts which we 
believe are instrumental effects and which we do not consider further
in this paper. There is no evidence of rotational modulation, and only one flare is
evident in the data set.  

\begin{figure*}
\plotone{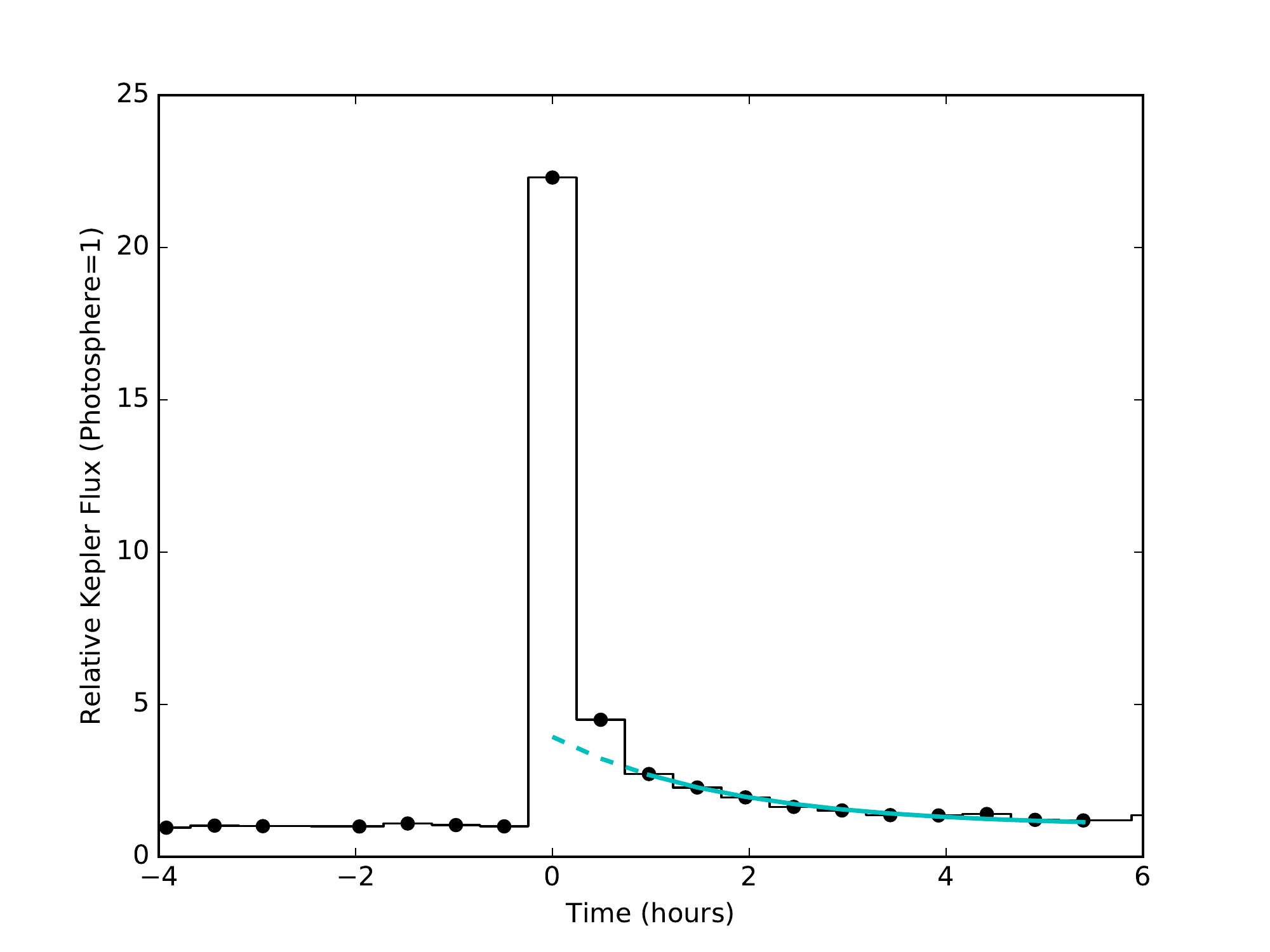}
\caption{K2 photometry of a flare on SDSSp J005406.55-003101.8. Time zero on this plot is {\it Kepler} mission day 2595.78419378.  The cyan curve shows our fit to the late-time exponential decay (Equation~\ref{eqn1}); the extrapolation back to the time of the flare is shown as the dashed curve. The count rate outside of the flare was 84.5 count s$^{-1}$.    
\label{fig1}}
\end{figure*}

At {\it Kepler} mission day 2595.784194, the target brightens to 1880 count s$^{-1}$ ($\widetilde{K}_p=17.1$, $\Delta \widetilde{K}_p = 3.4$); by the next 30 minute cadence, it faded to 380 count s$^{-1}$, and then continued to decline over the next several hours (Figure~\ref{fig1}). The fast rise and slow/exponential decline is typical of stellar flares in the Kepler band. \citet{2014ApJ...797..122D} described the Kepler light curve of an M4 flare star as double exponential, consisting of both an ``impulsive" decay component which dominates at early times and a ``gradual" decay component which dominates at late time. We adopt this terminology. The initial impulsive rise and decay is not resolved in our time series, implying the full time width at half the maximum flux ($t_{1/2}$, \citealt{2013ApJS..207...15K}) is less than 30 minutes, but a gradual decay component is clearly detected. \citet{2015EP&S...67...59M} define the ``duration" of the flare as the ``e-folding decay time of flare intensity after
its peak," which also must be less than 30 minutes; our fits discussed in the next section suggest the value of $\sim 7-10$ minutes.  \citet{2014ApJ...797..121H} define ``duration" instead as the difference between the start and end times, meaning the times when the flare flux is detectable above the photospheric flux. The duration under this definition is 5 hours. 

Because the ASASSN-16ae flare on S0533+00 was observed in the V-band, we estimate the V-band properties of our S0054$-$00 flare. At the peak of the flare, S0054$-$00 would have become a blue source rather than an extremely red one. Taking an A star photosphere as a proxy for the hot emission of a flare, we find that $V=K_p -0.1$ for A stars when we transform from $gri$ to V and $K_p$ \citep{2011AJ....142..112B,2016ApJS..224....2H},  implying that the observed peak of the flare was $V=17.0$ and therefore $\Delta V=-7.1$.  
The observed peak is the average over a 30 minute period which likely underestimates the true peak brightness of the flare but the integrated counts are correct. 
The total ``equivalent duration", which is related to the total energy of the flare and should not be confused with the duration previously discussed, is equal to the total integrated counts of the flare divided by the photosphere count rate \citep{1972Ap&SS..19...75G,2014ApJ...797..121H}, 15.4 hours in the {\it Kepler} filter. Equivalent duration is bandpass-dependent:  In V-band, the equivalent duration was $\sim17$ days.

A number of different approaches have been used to describe the energy of flares detected by Kepler.  
\citet{2011AJ....141...50W} and \citet{2014ApJ...792...67C} report bolometric (UV, visible and infrared, but not X-ray) energies.  The flare spectral energy distribution is extrapolated beyond the {\it Kepler} filter by assuming the flare is a blackbody of temperature 10,000K.
Using the L1 dwarf and 10,000K blackbody calibrations described by \citet{2013ApJ...779..172G}, we find a total flare energy of $E_{\rm bol} \approx 4\times10^{33}$ erg. 
This extrapolation to bolometric energy is uncertain to at least the $\sim20$\% level (see Table 3 of \citealt{2013ApJ...779..172G}). It is also useful to instead consider the observed energies through the {\it Kepler} filter. \citet{2014ApJ...797..121H} report ``Kepler energies"  ($E_K$) which are tied to the observed spectrum of the M4 flare dwarf GJ 1243 and its input catalog magnitude, and do not extrapolate to include the ultraviolet or infrared. These cannot be directly compared to our analysis because the L1 dwarf is much cooler than GJ 1243, but we can follow their procedure to compute $E_K$ using the S0054$-$00 spectrum.\footnote{Another difficulty is that $K_p$ system was not designed to predict M dwarf brightnesses: In the case of GJ 1243, although $K_p = 12.7$, we measure $\widetilde{K}_p = 11.65$ from the {\it Kepler} pixel data.} Convolving the S0054$-$00 photosphere spectrum with the {\it Kepler} response function, we find the stellar photospheric flux is $9.9 \times 10^{-18}$ erg s$^{-1}$ cm$^{-2}$ \AA$^{-1}$, but a 10,000K blackbody is 1.3 times more energetic for the same count rate. 
Following the same procedure as \citet{2014ApJ...797..121H}, we multiply by the width of the filter (4000\AA), $4 \pi d^2$, and the {\it Kepler} equivalent duration. After applying our energy correction factor of 1.3, we find $E_K = 9 \times10^{32}$ erg for the flare.  
    
\section{Template Fitting} \label{templates}

With a total estimated energy of $4 \times 10^{33}$ erg, the S0054$-$00 flare lies between the strongest flare observed on the L dwarf W1906+40 ($\sim1.6 \times 10^{32}$ erg, \citealt{2013ApJ...779..172G}; we hereafter call this ``the W1906+40 flare") and the strongest L dwarf flare, ASASSN-16ae ($>6.2\times10^{34}$ erg,\citealt{2016ApJ...828L..22S}). The ASSASN-16ae curve was only sparsely sampled; the advantage of the S0054$-$00 flare is that the full light curve, especially the gradual decay phase, was observed. This allows us to characterize the late cooling phases of such an energetic flare for the first time.  

We first consider the late-time, gradual decay phase of the flare. After normalizing by the non-flaring photosphere (84.5 count s$^{-1}$) and measuring from the time of the flare ($\Delta t=t-2595.78419378$), we fit\footnote{All fits reported in this paper are the result of Markov Chain Monte Carlo sampling of at least 1.5 million samples of the posterior function using {\rm emcee} \citep{2013PASP..125..306F} with uniform priors.} an exponential decay curve to the ten data points from 1 to 5.5 hours after the peak: 

\begin{equation}
\Delta F =  B e^{-\gamma_g \Delta t}
\label{eqn1}
\end{equation}

We find $B= 2.94\pm0.22$ and $-\gamma_g = -13.6 \pm 1.0$ days$^{-1}$; this fit is shown in Figure~\ref{fig1}. The exponential time constant ($\tau_g = 1/\gamma_g$ ) is 1.8 hr. \citet{2016ApJ...829..129S} analyzed
simultaneous spectroscopic and {\it Kepler} observations of the late exponential decay to conclude that in case of flares on the M4.5 dwarf GJ 1243, the gradual decay phase should be considered to be due to a ``physically distinct" region from the impulsive decay phase,  due to perhaps a ``different spatial region, different atmospheric layer, or different cooling process."  In the W1906+40 flare, which was observed spectroscopically, the Kepler impulsive and gradual decay phases traced the white light emission and broadened H$\alpha$, but atomic (chromospheric) emission lines were much longer lived.  

To investigate the impulsive phase of the flare we turn to model templates based on other white light flares to infer what may have happened during the first hour of the flare. Ground-based photometry of flares show a great diversity of light curves, but remarkably, \citet{2014ApJ...797..122D} found that a single flare template (hereafter the D14 template) fits most {\it Kepler} flares on the M4 dwarf GJ 1243, once the flares were scaled by their maximum amplitude ($A$) and full width at half time ($t_{1/2}$). The rapid rise is described by a fourth order polynomial, and their decay template has two exponentials, which we write as:

\begin{equation}
\Delta F =  A (\alpha_i e^{-\gamma_i \Delta t/t_{1/2}} + \alpha_g e^{-\gamma_g \Delta t/t_{1/2}})
\label{D14equation}
\end{equation}

\begin{figure*}
\plotone{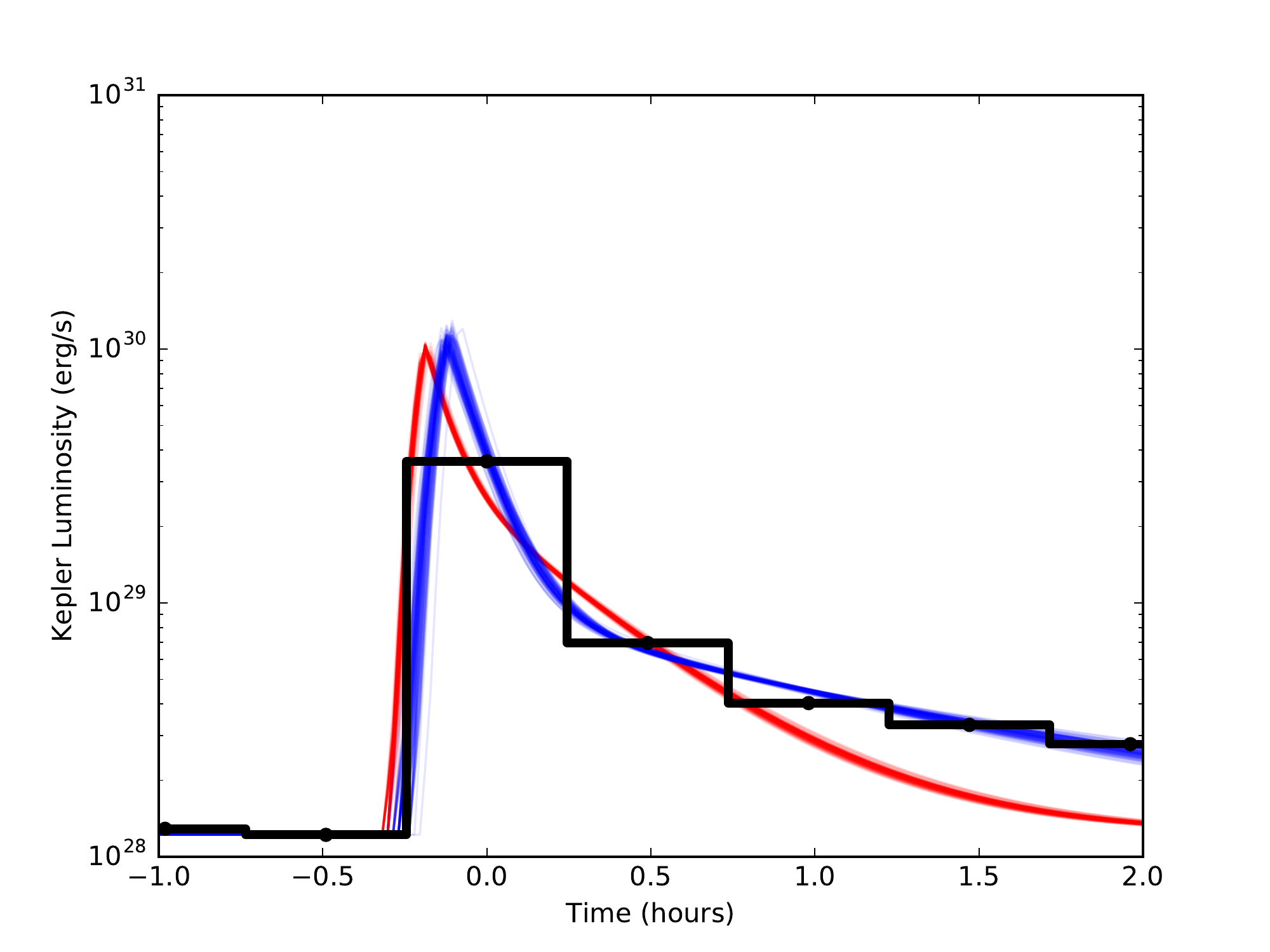}
\caption{K2 photometry with illustrative flare light curve fits. The original Davenport flare light curve template is shown in red, and an alternative template, based on a fit to an M7 brown dwarf flare, is shown in blue. In each class, 100 randomly chosen fits from the posterior distribution are plotted.   
These fits suggest that the flare amplitude would have been much higher ($\Delta K_p \approx -4.6$)  if sampled on one minute timescales instead of averaged over thirty minutes. See the text for further discussion. Time zero on this plot is mission time 2595.78419378.    
\label{fig2}}
\end{figure*}

The D14 template parameters are $\alpha_i=0.6890 (\pm 0.0008)$, $\alpha_g = 0.3030 (\pm 0.0009)$, $\gamma_i=1.600(\pm0.003)$ and $\gamma_g = 0.2783(\pm0.0007)$, so that along with the polynomial parameters which we do not list here, there are nine fixed parameters common to all flares, plus three free parameters ($A$, $t_{1/2}$ and time of flare) that are unique for each flare. This template also fits the W1906+40 flare, where we find $t_{1/2} = 6.9$ minutes, and was used by \citet{2016ApJ...828L..22S} to fit ASASSN-16ae to find $t_{1/2}$ in the range 3 (best fit) to 6.2 (minimal fit) minutes. We fit the D14 template to our flare by computing it on one minute cadence timescale and averaging down to the long cadence timescale. Because we are interested in the impulsive phase,  we fit to only the first 1.5 hours of the flare (plus two hours before the flare, for a total of seven data points). The results are shown in Figure~\ref{fig2}. One hundred randomly drawn model fits from the posterior distribution are shown in red. The three derived parameters are flare amplitude $A=63.9^{+1.0}_{-0.9}$, $t_{1/2} = 6.69 \pm 0.14$ minutes, and flare time $2595.77636 \pm 0.00017$ dy. It is clear from Figure~\ref{fig2} that the late time evolution is not well fit by this template. We can show this directly from our gradual phase fit: For the D14 template to reduce to Equation~\ref{eqn1} at late times, the maximum amplitude of the flare would be $2.94/0.303 = 9.7$ times the photospheric count rate and $t_{1/2}  = 0.2783/13.6$ day $=29.5$ min, but this amplitude is too low and the timescale is too long compared to the K2 observations. Unfortunately, the S0054$-$00 long cadence data does not contain enough content to allow all the possible flare parameters of Equation~\ref{eqn1} to be fit independently.
To illustrate alternative possibilities, we also make use of template parameters that were fit to a very impulsive flare on an M7 brown dwarf observed by K2 in short cadence mode (Gizis et al., in prep): $\alpha_i=0.9233$, $\alpha_g=0.0767$, $\gamma_i = 1.3722$, and $\gamma_g=0.1163$. Fixing those four template parameters plus the original D14 polynomial rise parameters, we fit $A=69.9\pm 5.5$, $t_{1/2} = 7.8^{+0.8}_{-0.7}$ min, and time of the flare $2595.77918 \pm 0.00050$ days to the S0054$-$00 flare. The light curve for this fit is a better match to the gradual decay phase (Figure~\ref{fig2}.) Although our lack of detailed, independent knowledge of the impulsive flare light curve means we cannot be precise on the true amplitude and timescale of the S0054$-$00 flare, the template fitting exercise suggests that the timescale ($t_{1/2}$) is less than $10$ minutes and that the true flare maximum may have been approximately seventy times the mean photospheric value ($\Delta \widetilde{K}_p \approx -4.6$) implying $\Delta V \approx -8$, more than three times that observed in the data averaged over thirty minutes.   
 
\section{Discussion}\label{discussion}

All three L dwarfs with observed white light flares appear to be old, hydrogen-burning stars with ages measured in the billions of years; indeed S0533+00 (ASASSN-16ae) is likely a thick-disk star, perhaps ten billion years old. White light flares and superflares may be common to all early-L dwarfs regardless of age. Evidently the rapid rotation and weak spin-down of this class of stars supports a dynamo that generates robust magnetic fields and reconnection events that can exceed solar flare energies.  \citet{2015EP&S...67...59M} calculate superflares of energy $10^{33}$ erg occur about once every seventy years on the Sun, which contrasts with the one superflare observed in $1/5$ of a year of monitoring S0054$-$00 with K2.  However, a more realistic, and lower, flare rate will come by considering all ($\sim20$) of the early-L dwarfs monitored during the full K2 mission. Long-term monitoring of S0054$-$00 may reveal additional flares --- we note that the Catalina Real-Time Transient Survey \citep{2009ApJ...696..870D} 
found S0054$-$00 to be approximately one magnitude brighter than normal in January 2006 (MJD 53744.14) in unfiltered CCD images. This may have been a white light flare. \citet{2016ApJ...828L..22S} use the ASASSN data to estimate a flare rate of approximately one in six years for L0 dwarf flares with energy $10^{34}$ erg.

For these three flaring L dwarfs, the flare duration as measured by the e-folding timescale is $<10$ min for energies $10^{33}-10^{34}$ erg, which agrees with the observed duration of G dwarf superflares  \citep{2015EP&S...67...59M} with the same energy despite the much different stellar size and effective temperatures. Given the similarity of the L dwarf flares to G dwarf flares, it is interesting to compare them to the solar flare simulations of \citet{2013A&A...549A..66A} which can account for flares and superflares in G dwarfs, although we caution that physical conditions may be different. They find the following scaling of bolometric energy with maximum magnetic field strength and linear separation between the bipoles ($L_{\rm dipole}$) :

\begin{equation}
E = 0.5 \times 10^{32} \left({B_z^{\rm max}}\over{1000G}\right)^2 \left({L^{\rm bipole}}\over{50 {\rm Mm}}\right)^3 {\rm erg}
\end{equation}.

If we use the radius of the L1 dwarf ($0.09 R_\odot$, \citealt{2014AJ....147...94D}) as the length scale, corresponding to a pair of large starspots (active regions) separated by $\sim60^\circ$, then maximum magnetic field strengths of 6000G are needed for the S0054$-$00 superflare. \citet{2006ApJ...648..629B} report magnetic fields of $\sim1000$G in much less energetic ultracool dwarf radio flares, so such a field in an unusually strong flare seems plausible. For ASASSN-16ae, increasing the volume by a factor of 10 (such as with spots on the opposite sites of the star) and increasing the field to 8000G could account for the observed energy. While the lengths needed to produce superflares with $E>10^{35}$ erg may be possible in solar-type stars, the small sizes of single L dwarfs suggest that they will not have higher energy superflares unless extremely strong ($>10$kG) magnetic fields are possible. 

L dwarf superflares offer significant opportunities to study the later phases of a superflare, particularly the gradual cooling phase. In solar-type stars, the bright photosphere makes even the peak emission at best a few percent effect, and following the late decay of a superflare in white light would require extraordinary precision. For L dwarfs, however, the photospheric contribution is greatly reduced. In the flare presented here, a trigger from a photometric survey like LSST with 30 minute response time on a 8-meter class telescope would allow hours of spectroscopy of both the gradual phase white light component and atomic emission lines. Such observations will yield new constraints on the cooling of stellar photospheres in response to impulsive energy injections.  

\acknowledgments

We thank James Davenport, Dermott Mullan, and Rachel Osten for helpful discussion of stellar flares.  
This paper includes data collected by the Kepler mission. Funding for the Kepler mission is provided by the NASA Science Mission directorate. The material is based upon work supported by NASA under award Nos. NNX15AV64G, NNX16AE55G, and NNX16AJ22G.  A.J.B. acknowledges funding support from the National Science Foundation under award No.\ AST-1517177. 

This research has made use of NASA's Astrophysics Data System, 
the SIMBAD database and he VizieR catalogue access tool, operated at CDS, Strasbourg, France, 
the NASA/ IPAC Infrared Science Archive, which is operated by the Jet Propulsion Laboratory, California Institute of Technology, under contract with NASA, and the Mikulski Archive for Space Telescopes (MAST). Support for MAST for non-HST data is provided by the NASA Office of Space Science via grant NNX09AF08G and by other grants and contracts. This publication makes use of data products from the Two Micron All Sky Survey, which is a joint project of the University of Massachusetts and the Infrared Processing and Analysis Center/California Institute of Technology, funded by the National Aeronautics and Space Administration and the National Science Foundation and from the Wide-field Infrared Survey Explorer, which is a joint project of the University of California, Los Angeles, and the Jet Propulsion Laboratory/California Institute of Technology, funded by the National Aeronautics and Space Administration. This publication also makes use of data from the Sloan Digital Sky Survey. Funding for the Sloan Digital Sky Survey IV has been provided by the Alfred P. Sloan Foundation, the U.S. Department of Energy Office of Science, and the Participating Institutions. SDSS-IV acknowledges support and resources from the Center for High-Performance Computing at the University of Utah. The SDSS web site is \url{www.sdss.org}. SDSS-IV is managed by the Astrophysical Research Consortium for the Participating Institutions of the SDSS Collaboration, which can be found at \url{http://www.sdss.org/collaboration/affiliations/}

\software{AstroPy \citep{2013A&A...558A..33A}, photutils, emcee \citep{2013PASP..125..306F}}

\facility{Kepler}

\bibliographystyle{aasjournal}
\bibliography{../astrobib}


\end{document}